\documentclass[showkeys,preprint,english,11pt]{revtex4}
\usepackage{amsmath}
\usepackage{esint}
\usepackage{amssymb}
\usepackage{graphicx, epsfig}
\usepackage{mathrsfs}
\usepackage{hyperref}
\usepackage{bm}
\usepackage {amsmath}
\makeatletter
\usepackage{babel}
\makeatother
\allowdisplaybreaks

\begin{document}

\title{\bf{A Dark Energy Model in Lyra Manifold}}

\author{Hoavo Hova\footnote{hovhoav@mail.ustc.edu.cn}}

\affiliation{Interdisciplinary Center for Theoretical Study, University of
Science and Technology of China, Hefei, 230026, P. R. China}

\date{\today}

\begin{abstract}
We consider, in normal-gauge Lyra's geometry, evolution of a homogeneous isotropic universe in a gravitational model involving only the standard matter in interaction with a displacement vector field $\phi_{\mu}$. Considering both constant and time-dependent displacement vector field we show that the observed cosmic acceleration could be explained without considering an alien energy component with a negative pressure.
\end{abstract}

\maketitle

\section{Introduction}
There is increasing evidence from recent cosmic observations of  Type Ia Supernovae\cite{Riess, Perlmutter}, Large Scale Structure (LSS)\cite{Tegmark1, Tegmark2, Tegmark3, Seljak, Adelman, Abazajian} and Cosmic Microwave Background(CMB)\cite{Spergel1, Page, Hinshaw, Jarosik} that the universe is undergoing an accelerated expansion at the present stage. This phenomenon indicates that the universe at present is dominated by a smooth energy component,  dubbed ``dark energy'', with a negative pressure that counteracts the gravitational forces produced by ordinary matter species, such as baryons and radiation, leading then to an accelerated expansion of the universe. Despite many years of research and much progress, the nature and the origin of dark energy have not been confirmed yet.

Obviously, the best and simplest candidate for such dark energy is the so-called cosmological constant (CC) $\Lambda$ which was introduced by Einstein into his gravitational field equations in an \emph{ad hoc} fashion. However, CC explanation for dark energy usually faces some fundamental problems in physics, namely the fine-tuning problem and notably the cosmic coincidence puzzles \cite{Weinberg, Copeland, Amendola1}. In particle physics, the CC is often interpreted as the energy, $\rho_{vac}$, of the quantum vacuum, which is close to the Planck density $M_{P}^{4}$ ($M_{P}=1/\sqrt{8\pi G}$ is the reduced Planck mass) in magnitude. The observed value of the dark energy density is much less than that of the quantum vacuum, $\rho_{obs} \approx 10^{-123}\rho_{vac}$. Suppressing this great difference of about 123 orders of magnitude between the observed value of dark energy and that estimated from quantum field theory requires some severe fine-tuning mechanisms to work\cite{Hawking, Kachru, Tye, Yokoyama, Mukohyama, Kane, Dolgov}. On the other hand, even if this fine-tuning problem could be evaded, the non-dynamical behaviour of the quantum vacuum energy renders the coincidence problem unsolved. Attempting to alleviate these two fundamental problems and to explain the late-time cosmic acceleration, many plausible dynamical models, such as quintessence \cite{Ratra}, phantom \cite{Onemli, Caldwell}, k-essence \cite{Armendariz}, tachyon \cite{Padmanabhan}, holographic \cite{Li, Limiao}, agegraphic \cite{Cai}, hessence \cite{Wei}, Chaplygin gas \cite{Kamenshchik}, Yang-Mills condensate \cite{Zhang}, etc., have been proposed (see also Review article \cite{Copeland} and references therein). 

A different approach to explain the observed accelerating universe with models involving only the standard matter is a plausible modification of the Einstein gravitational field equations . Such a modification can arise, either by extending the Einstein-Hilbert action to a more fundamental theory ($f(R)$ theories of gravity  \cite{Kerner, Capozziello, Nojiri, Capozziello1, Nojiri1}) or by modifying the Riemannian geometry. In the latter case, a Lyra's geometry \cite{Lyra, Sen1, Sen2}, which bears a close similarity to  Weyl's geometry \cite{Weyl} and is propounded in order to unify gravitation and electromagnetism into a single space-time geometry, got lots of interest. Indeed, in contrast to Weyl's geometry,  the connection in Lyra's geometry is metric preserving, as in Riemannian geometry, and length transfers are also integrable. In addition, theories of gravitation, that have been constructed in the framework of Lyra's geometry with both a constant and a time-dependent displacement vector field,  involve  scalar fields and tensors that are all intrinsic to the geometry \cite{Sen1, Sen2, Sen3, Sen4, Scheibe, Halford, Halford1, Soleng, Soleng1, Hudgin, Beesham, Beesham1, Manoukian, Matyjasek, Anirudh, Mohanty, Rahaman, Katore, Gad, Shchigolev}. On the other hand, as shown in \cite{Sen1, Halford1} these theories predict the same effects within observations limits, as far as the classical Solar System, as well as tests based on the linearised form of the field equations, and are free of the Big-Bang singularity and solve the entropy and horizon problems, which beset the standard models based on Riemannian geometry.

In the present work, we consider a pressureless matter in interaction with the displacement vector field. As pointed out in \cite{Shchigolev}, we will see that, in the absence of a pressureless matter the displacement vector field alone could not be considered as a cosmological constant (term) but rather a stiff fluid, because the associated equation of state is $\omega_{\phi}=+1$ and not $\omega_{\textsc{cc}}=-1$. Meanwhile, interacting with the pressureless matter, the displacement vector can play the same role as a cosmological constant (term), establishing therefore the intrinsic geometrical origin of the cosmological term. Subsequently, it is shown that the observed  accelerating universe can occur without considering an alien energy component with a negative pressure.

The outline of the paper is as follows.  In Sect. II, firstly we derive the $\Lambda$CDM model from a model containing the standard matter and a constant displacement vector field, and secondly we focus on a time-dependent vector field that yields a variable cosmological term and drives then an accelerating universe.  A summary of the results and the conclusions are presented in Sect. III. Throughout the paper we adopt the Planck units $c=\hbar=\kappa^{2}=1$ and use the \textit{space-like metric signature} $(-,+,+,+)$.

\section{Cosmic Acceleration in Normal-Gauge Lyra manifold} 
The Einstein gravitational field equations in normal gauge for a four-dimensional Lyra manifold, as obtained by Sen \cite{Sen1} are
\begin{equation}
\label{h}
G_{\mu\nu}=T_{\mu\nu}+{\cal T}_{\mu\nu},
\end{equation}
where 
\begin{equation}
{\cal T}_{\mu\nu}=-\frac{3}{2}\left(\phi_{\mu}\phi_{\nu}-\dfrac{1}{2}g_{\mu\nu}\phi_{\lambda}\phi^{\lambda} \right),
\end{equation}
is the stress-energy tensor associated with the displacement vector field $\phi_{\mu}$, i.e., arising as an intrinsic geometrical energy-momentum tensor, whereas $T_{\mu\nu}$ represents a perfect fluid energy-momentum tensor defined by
\begin{equation}
T_{\mu\nu}=\left( p+\rho\right)u_{\mu}u_{\nu} +pg_{\mu\nu}.
\end{equation}
 $u_{\mu}=(1,0,0,0)$ is the 4-velocity of the comoving observer, satisfying $u^{\mu}u_{\mu}=-1$, $\rho$ and $p=\omega_{b}\rho$ with $0\le \omega_{b}< 1$  are the background energy density and pressure, respectively. In this paper we shall consider a homogeneous time-depending time-like displacement vector 
\begin{equation}
\phi_{\mu}=\left(\phi(t),0,0,0 \right)
\end{equation}
 With these assumptions the $(0,0) $ and $(i,j)$-components of Eq. (\ref{h}), in the flat \textit{Friedmann-Lema\^{i}tre-Robertson-Walker} background $ds^{2}=-dt^2+a^{2}d\vec{x}^{2}$, where $a=a(t)$ is the scale factor of an expanding universe, may be written
\begin{eqnarray}
\label{2}
3H^{2}=\rho-\frac{3}{4}\phi^{2}(t)=\rho_{eff},\\
\label{3}
-\left( 2\dot{H}+3H^{2}\right) =\omega_{b}\rho-\frac{3}{4}\phi^{2}(t)=p_{eff},
\end{eqnarray}
where an overdot denotes differentiation with respect to the time coordinate $t$ and $H=\dot{a}/a$ is the Hubble parameter. In the above equations we can recast the pressure and energy density of the displacement vector in the forms $p_{\phi}=\rho_{\phi}=-\frac{3}{4}\phi^{2}(t)$, giving a constant equation of state, $\omega_{\phi}=+1$,  of a stiff matter. This proves that the displacement vector alone cannot be considered as a cosmological constant which, by contrast, has an equation of state $\omega_{\textsc{cc}}=-1$. In the following, we set $\theta\equiv \phi^{2}$ and refer to $\theta$ as \textit{displacement field}. Thus, arguing the displacement field is a geometrical energy component contributing to the total energy and interacting with the standard matter, we will show that the displacement vector can play the role of a cosmological constant (term). In this context we consider both a constant and time-dependent displacement vector field.

\subsection{Constant Vector Field and Cosmological Constant}
Considering a constant vector field $\phi^{2}=\phi^{2}_{0}=\theta_{0}$ and starting with a background fluid with a constant equation of state $\omega_{b}$ the Friedmann equations translate into
\begin{eqnarray}
\label{2.1}
3H^{2}=\rho-\frac{3}{4}\theta_{0}=\rho_{eff},\\
\label{2.2}
2\dot{H}+3H^{2}=-\omega_{b}\rho+\frac{3}{4}\theta_{0}=-p_{eff}.
\end{eqnarray}
Resolving the continuity equation of the effective energy density, $d\rho_{eff}/dx+3(\rho_{eff}+p_{eff})=0$ where $x=\ln a$ is the e-folding number,  one obtains
\begin{eqnarray}
\label{11}
\rho=ke^{-3(1+\omega_{b})x} +\frac{3}{2(1+\omega_{b})}\theta_{0},\\
\label{11a}
3H^{2}=\rho_{eff}=ke^{-3(1+\omega_{b})x} +\frac{3}{4}\dfrac{1-\omega_{b}}{1+\omega_{b}}\theta_{0},\\
\label{11b}
p_{eff}=\omega_{b}ke^{-3(1+\omega_{b})x}-\frac{3}{4}\dfrac{1-\omega_{b}}{1+\omega_{b}}\theta_{0},
\end{eqnarray}
whence we can compute the effective equation of state as follows
\begin{equation}
\omega_{eff}=-1+\dfrac{1+\omega_{b}}{1+n_{0}e^{3(1+\omega_{b})x}}
\end{equation}
with 
\begin{equation}
n_{0}=\frac{3}{4}\dfrac{1-\omega_{b}}{1+\omega_{b}}\dfrac{\theta_{0}}{k}.
\end{equation}
Hence the effective equation of state varies from $\omega_{b}$ at $x\to-\infty$ to $\omega_{\textsc{cc}}=-1$ at $x\to+\infty$.

Considering now a pressureless matter $\omega_{b}=0$, Eqs. (\ref{11})-(\ref{11b}) reduce to the $\Lambda$CDM-like model
\begin{eqnarray}
\label{matter}
\rho=ke^{-3x} +\frac{3}{2}\theta_{0},\\
\label{2.3}
3H^{2}=\rho_{eff}=ke^{-3x} +\frac{3}{4}\theta_{0},\\
\label{press}
p_{eff}=-\frac{3}{4}\theta_{0}.
\end{eqnarray}
From Eq. (\ref{matter}) we say that the pressureless background fluid $\rho$ is the contribution of two terms: the cold dark matter (CDM) and a gain of energy $\varepsilon=\frac{3}{2}\theta_{0}$  from a modification of the Riemannian manifold by the presence of a vector field in the geometrically structureless manifold. On the other hand, Eq. (\ref{2.3}) shows that the effective energy density is also a sum of two terms, namely   the CDM and a contribution from the displacement field, $\tilde{\epsilon}=\frac{3}{4}\theta_{0}$. Since the effective pressure (\ref{press}) is the contribution of the displacement field only, one finds that through the conservation of the total energy density the equation of state for the displacement field becomes $\omega_{\phi_{int}}=p_{eff}/\tilde{\epsilon}=-1$. In this case the displacement field (vector) plays the same role as the cosmological constant $\Lambda$. 

The effective equation of state and the fractional Hubble parameter $E\equiv H/H_{0}$ are therefore given by
\begin{equation}
\omega_{eff}=-\frac{1}{1+\lambda(1+z)^{3}},
\end{equation} 
\begin{equation}
E^{2}(z)=\Omega_{m}(1+z)^{3}+(1-\Omega_{m}),
\end{equation}
where $z=a^{-1}-1$ is the redshift, $\lambda=\frac{4k}{3\theta_{0}}=\dfrac{\Omega_{m}}{1-\Omega_{m}}$, $\Omega_{m}\equiv \dfrac{k}{3H_{0}^{2}}$ is the matter density parameter. Thus, at present time ($x=0$) acceleration occurs for $\lambda<2$  or $\Omega_{m}<2/3$. A model involving a pressureless background fluid in normal gauge for Lyra manifold is therefore equivalent to the $\Lambda$CDM model if one sets $\Lambda=\frac{3}{4}\theta_{0}$. We conclude that this model shows the intrinsic geometrical origin of the cosmological constant with $\omega_{\textsc{cc}}=\omega_{\phi_{int}}=-1$, and the constant displacement vector field arises therefore as the origin of the late time accelerated expansion of the universe.

\subsection{Time-Dependent Vector Field}
The vector field $\phi_{\mu}$ now depending on time interacts mutually with the background fluid $\rho$. And the effective energy conservation equation can be written as
\begin{equation}
\label{2.5}
\frac{d\rho}{dx}+3(1+\omega_{b})\rho-\frac{3}{4}\left(\frac{d\theta}{dx}+6\theta \right)=0, 
\end{equation}
 Eq. (\ref{2.5}) involves two unknown functions $\rho$ and $\theta$, that means in addition to the Hubble parameter we have three unknown functions to be determined with only two independent equations.  In what follows we construct the cosmological consequences of these equations, under some conditions that significantly simplify the search for solutions, but nevertheless show the richness of the cosmological dynamics of the present model. We can thus encode the interaction between $\rho$ and $\theta$ into the conservation equations \cite{Amendola1, Limiao}
\begin{eqnarray}
\label{2.7}
\frac{d\rho}{dx}+3(1+\omega_{b})\rho=\gamma(x),\\
\label{2.8}
\frac{3}{4}\left(\frac{d\theta}{dx}+6\theta \right)=\gamma(x),
\end{eqnarray}
where $\gamma$ denotes the phenomenological interaction term.  Owing to the lack of the knowledge of micro-origin of the interaction and following other works \cite{Limiao} we simply parametrize the interaction term in the form: 
\begin{equation}
\label{2.9}
\gamma=3b\rho+\frac{9}{2}\tilde{b}\theta~,
\end{equation}
where both $b$ and $\tilde{b}$ are dimensionless coupling constants, and the factors $3$ and $9/2$ before $b$ and $\tilde{b}$ are for convenience in the following calculations. We emphasize here that $\gamma$ does not contain explicitly the Hubble parameter $H$, because we are using the e-folding number $x=\ln a$. However, once we are dealing with the cosmological time $t$, we instead use $\tilde{\gamma}=H\gamma=\Gamma\rho+\tilde{\Gamma}\theta$, where $\Gamma$ and $\tilde{\Gamma}$ characterize the strength of the coupling. Without loss of generality and in order to reduce, furthermore, the number of parameters we shall consider in this section a pressureless matter and then set $\omega_{b}=0$. In the following we shall study the cases: $i)$ $b\ne 0$ and $\tilde{b}=0$, $ii)$ $b=0$ and $\tilde{b}\ne 0$ and $iii)$ $b\ne 0$ and $\tilde{b}\ne 0$, respectively.

\subsubsection*{Case $i$:~~ $b\ne 0$ and $\tilde{b}=0$} 
Solving (\ref{2.7}) and (\ref{2.8}) we find
\begin{eqnarray}
\theta= -M_{1}e^{-6x}+\frac{4bC_{1}}{3(1+b)}e^{-3(1-b)x},\\
\rho= C_{1}e^{-3(1-b)x},\\
\label{2.11}
3H^{2}=\rho_{eff}=\frac{C_{1}}{1+b}e^{-3(1-b)x}+\frac{3 M_{1}}{4}e^{-6x},\\
\label{2.12}
p_{eff}=\frac{3 M_{1}}{4}e^{-6x}-\frac{b C_{1}}{1+b}e^{-3(1-b)x},
\end{eqnarray}
where $C_{1}$ and $M_{1}$ are integration constants. We realize that, with this choice of $\gamma$ and for $M_{1}>0$, the pressureless fluid energy and the effective energy density are always positive and monotonically decreasing for  $C_{1}>0$ and $-1<b<1$ during the evolution of the universe. Meanwhile the effective pressure could be either positive in early times (the pressureless matter dominates over the displacement field) or negative at the present stage (vector field dominance epoch). We can now evaluate the dynamically varying effective equation 
\begin{equation}
\omega_{eff}=\frac{1-b\xi e^{3(1+b)x}}{1+\xi e^{3(1+b)x}}~,
\end{equation} 
where $\xi=\frac{4C_{1}}{3M_{1}(1+b)}$ is a positive dimensionless constant for $b>-1$. Assuming a monotonically decreasing  effective energy density and positive $\xi$,  the coupling constant $b$ needs to range in the region $-1<b<1$. The effective equation of state then lies in the range $\omega_{eff}^{+\infty}=-b<\omega_{eff}<\omega_{eff}^{-\infty}=1$. At the present stage we have
\begin{equation}
\omega^{0}_{eff}=-b+\dfrac{1+b}{1+\xi},
\end{equation}
which is less than $\omega_{acc}=-1/3$ for $\xi>\dfrac{4}{3b-1}$. For example, $\omega^{0}_{eff}=-0.9$ if $\xi=\dfrac{1.9}{b-0.9}$. 

\subsubsection*{Case $ii$:~~ $b=0$ and $\tilde{b}\ne 0$} 
\begin{figure}[htb]
\begin{center}
\includegraphics[width=5in,height=3in,angle=0]{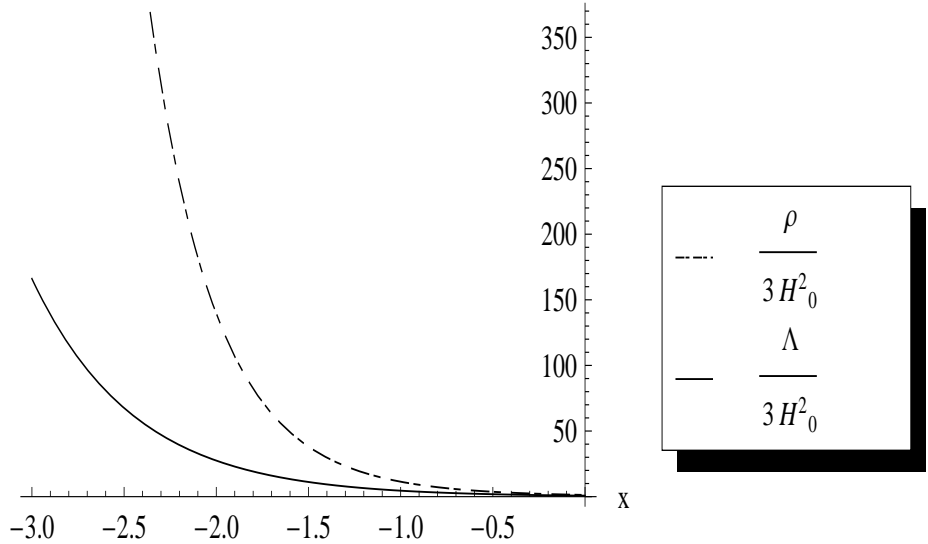} 
\caption[\emph{Evolution of $\rho/3H^{2}_{0}$ and $\Lambda/3H^{2}_{0}$ versus the e-folding number $x$}]{Evolution of $\rho/3H^{2}_{0}$ (dash-dotted curve) and $\Lambda/3H^{2}_{0}$ (solid curve) versus the e-folding number $x$ for $\varpi_{1}=0.3$ and $\tilde{b}=0.7$. In early times the cosmological term was negligible, compared to the pressureless energy density, while at the present stage they are of the same order. Hence, this model seems to solve both the fine-tuning and the coincidence problems.}
\label{fig30.eps}
\end{center}
\end{figure}

In this case the different quantities, previously considered, become
\begin{eqnarray}
\theta=M_{2}e^{-6(1-\tilde{b})x},\\
\label{rho}
\rho=C_{2}e^{-3x}+\frac{3\tilde{b}M_{2}}{2(2\tilde{b}-1)}e^{-6(1-\tilde{b})x},\\
3H^{2}=\rho_{eff}=C_{2}e^{-3x}+\frac{3M_{2}}{4(2\tilde{b}-1)}e^{-6(1-\tilde{b})x},\\
\label{2.16}
p_{eff}=-\frac{3M_{2}}{4}e^{-6(1-\tilde{b})x},
\end{eqnarray} 
where $C_{2}$ and $M_{2}$ are integration constants. For $\tilde{b}>1/2$ both effective energy and pressureless background energy densities are positive-defined. The former is the sum of two terms: CDM and a dynamically variable quantity, $\Lambda(x)$, that could be thought of as  a \textit{variable cosmological-type term} associated with the vector field: 
\begin{equation}
\label{2.17}
\Lambda(x)=\frac{3\varpi_{1}H_{0}^{2}}{2\tilde{b}-1}e^{-6(1-\tilde{b})x}~~\mbox{with}~~ \varpi_{1}=\dfrac{M_{2}}{4H_{0}^{2}}>0.
\end{equation}
 For $\tilde{b}<1$ the cosmological term is large during the early stages of the universe and has decayed to its low value at present, explaining readily the fine-tuning and coincidence puzzles (see fig.(\ref{fig30.eps}) where we plot both the pressureless energy density $\rho$ and the cosmological term $\Lambda$). On the other hand, the pressureless background fluid is also a sum of two terms, corresponding to CDM energy and an energy, $\rho_{geo-gained}$, geometrically gained as the universe evolves, that is,
\begin{equation}
\rho=\rho_{\textsc{cdm}}+\rho_{geo-gained},
\end{equation}
where 
\begin{equation}
\rho_{geo-gained}=\frac{6\varpi_{1}H_{0}^{2}}{2\tilde{b}-1}e^{-6(1-\tilde{b})x}
\end{equation}
is proportional to the cosmological term $\Lambda(x)$.

Since the effective pressure is the sum of the CDM pressure, $p_{\textsc{cdm}}=0$ and that of the variable cosmological term $\Lambda(x)$, from ({\ref{2.16}}) and ({\ref{2.17}}) one can determine the equation of state for the variable cosmological term: \[\omega_{\Lambda}\equiv\frac{p_{eff}}{\Lambda},\]\\
 or
\begin{equation}
\omega_{\Lambda}=1-2\tilde{b},
\end{equation}
 which, for $\tilde{b}\neq 1$, is different from the equation of state of the usual cosmological constant $\omega_{\textsc{cc}}=-1$. For example, for $\tilde{b}=0.9$, $\omega_{\Lambda}=-0.8$, and for $\tilde{b}=0.99$, $\omega_{\Lambda}=-0.98$. However, the choice $\tilde{b}>1$ will lead to dark energy behaving as phantom, i.e.,  $\omega_{\Lambda}$ being less than $-1$ \cite{Onemli, Caldwell, Kunz, Alam}, and  the variable cosmological term will increase exponentially from $0$ at $x\to -\infty$ to $+\infty$ at $x+\infty$. The ratio of the effective pressure over the effective energy density furnishes the effective equation of state in the form
\begin{equation}
\omega_{eff}=\frac{\omega_{\Lambda}}{1-\omega_{\Lambda}f_{0}e^{3\omega_{\Lambda}x}},
\end{equation} 
where 
\begin{eqnarray}
f_{0}&=&\frac{4C_{2}}{3M_{2}}\nonumber\\
&=&\dfrac{1}{\varpi_{1}}+\dfrac{1}{\omega_{\Lambda}}
\end{eqnarray}
 is supposed to be a positive constant, unlike $\omega_{\Lambda}$ which is always assumed negative. One thus has
\begin{equation}
\omega^{-\infty}_{eff}=0,~~~~ \omega^{0}_{eff}=-\varpi_{1},~~~\omega^{+\infty}_{eff}=\omega_{\Lambda}.
\end{equation}
As we have pointed out, a positive $f_{0}$ implies $\omega^{+\infty}_{eff}<\omega^{0}_{eff}$. Hence, $\varpi_{1}>1/3$ can drive an accelerated expansion. For example, taking $\varpi_{1}$ as the present fractional density of dark energy, $\omega^{0}_{eff}\approx -0.73$.

\subsubsection*{Case $iii$:~~ $b\ne 0$ and $\tilde{b}\ne 0$}
 Combining (\ref{2.7}) and (\ref{2.8}) and after some algebra it is shown that the effective energy density $\rho_{eff}$, the background fluid energy  $\rho$ and the displacement field  $\theta$ satisfy the same ordinary differential equation of second order under the form
\begin{equation}
\frac{d^{2}Q}{dx^{2}}+3(3-b-2\tilde{b})\frac{dQ}{dx}+18(1-b-\tilde{b} )Q=0,
\end{equation}
where $Q\in \left\lbrace \rho; \theta; \rho_{eff}\right\rbrace $. We will now consider some special cases where $b$ is explicitly related to $\tilde{b}$.
\begin{description}
\item[A/]~~$\tilde{b}=1-b$ ($b\ne 1$ or $\tilde{b}\ne 0$ ), one then obtains
\begin{equation}
\rho_{eff}(x)=N_{1}+\dfrac{N_{2}}{1+b}e^{-3(1+b)x},
\end{equation}
\end{description}
and the effective equation of state in this case takes for $b>-1$ the values $\omega_{eff}^{-\infty}=b$,~ $\omega_{eff}^{0}=-1+\varpi$ and $\omega_{eff}^{+\infty}=\omega_{\textsc{cc}}=-1$, where $\varpi=N_{2}/3H_{0}^{2}$. Acceleration thus occurs if $\varpi<2/3$. 
~~~~~~~~~~~~~~~~~~~~~~~~~~~~~\\
\begin{description}
\item[B/]~~$\tilde{b}=\dfrac{3-b}{2}$ ($b\ne 3$ or $\tilde{b}\ne 0$ ), $\rho_{eff}(x)$ takes the form
\end{description}
\begin{equation}
\rho_{eff}(x)=N_{3}e^{3\sqrt{1+b}~x}+N_{4}e^{-3\sqrt{1+b}~x},
\end{equation}
$N_{3}$ and $N_{4}$ being constants and $b>-1$. For monotonically decreasing $\rho_{eff}(x)$ as suggested observations we will set $N_{3}=0$, hence the effective equation of state is a constant, given by $\omega_{eff}=-1+\sqrt{1+b}$, which is less than $\omega_{acc}=-1/3$ and greater than $\omega_{\textsc{cc}}=-1$ for $-1<b<-5/9$.\\
~~~~~~~~~~~~~~~~~~~~~~~~~~\\
On the other hand, considering a nonzero $N_{3}$ but setting instead $N_{3}=N_{4}=r/2$, the effective energy density transforms according to
\begin{equation}
\label{spec}
\rho_{eff}(x)=r\cosh\left(3\sqrt{1+b}~x \right)  
\end{equation}
and
\begin{equation}
\label{spec1}
\omega_{eff}=-1-\sqrt{1+b}\tanh\left(3\sqrt{1+b}~x \right).
\end{equation}
The effective equation of state $\omega_{eff}$ (\ref{spec1}) varies from $\omega^{-\infty}_{eff}=-1+\sqrt{1+b}$, crosses $\omega_{\textsc{cc}}=-1$ at $x=0$ and tends to $\omega^{+\infty}_{eff}=-1-\sqrt{1+b}$. That thus leads to \emph{quintom} dark energy  with the equation of state crossing the cosmological constant boundary $\omega_{\textsc{cc}}=-1$ \cite{Guo}.\\

\begin{figure}[htb]
\begin{center}
\includegraphics[width=5in,height=3in,angle=0]{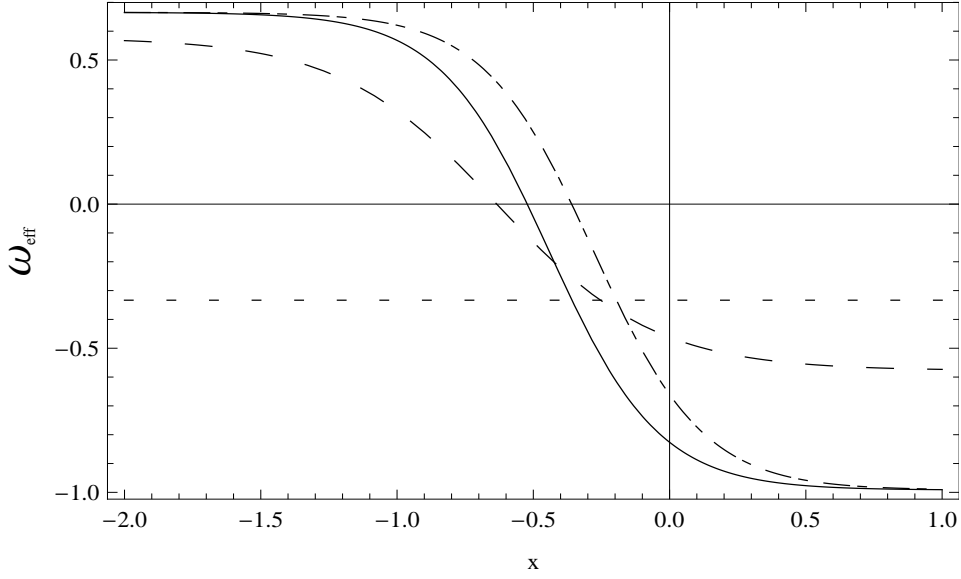} 
\caption[\emph{$\omega_{eff}$ versus x}]{$\omega_{eff}$ (\ref{specc}) versus $x$ for  $b=1/3,~ \tilde{b}=1/3,~ \chi=9$ (dashed curve, $\omega_{eff}\approx -0.46$);~ $b=0.66,~ \tilde{b}=1/3,~ \chi=4$ (dash-dotted curve, $\omega_{eff}\approx -0.66$) and $b=0.66,~ \tilde{b}=1/3,~ \chi=9$ (solid curve, $\omega_{eff}\approx -0.83$), the dotted line represents $\omega_{acc}=-1/3$.}
\label{fig31.eps}
\end{center}
\end{figure}

\begin{description}
\item[C/]~~Now, considering the cases where $\tilde{b}\ne 1-b$  and $\tilde{b}\ne\dfrac{3-b}{2}$  the general solution is given by
\begin{equation}
Q(x)=Q_{1}\exp\left[ -\frac{3}{2}\left(3-b-2\tilde{b}+\sqrt{\zeta} \right)x\right] 
+\tilde{Q}_{1}\exp\left[ -\frac{3}{2}\left(3-b-2\tilde{b}-\sqrt{\zeta}\right)x\right], 
\end{equation}
\end{description}
where  $Q_{1}\in \left\lbrace \rho_{1}; \theta_{1}; \rho^{1}_{eff}\right\rbrace $ and $\tilde{Q}_{1}\in \left\lbrace \tilde{\rho}_{1}; \tilde{\theta}_{1}; \tilde{\rho}^{1}_{eff}\right\rbrace $ are integration constants, and
\begin{equation}
\zeta=(3-b-2\tilde{b})^{2}-8(1-b-\tilde{b}).
\end{equation}
 The effective energy density reads then:
\begin{subequations}
\begin{eqnarray}
\label{special}
3H^{2}=\rho_{eff}&=&Q_{2}\exp\left[ -\frac{3}{2}\left(3-b-2\tilde{b}+\sqrt{\zeta} \right)x\right] 
~~~~~~~~~~~~~~~~~~~~~~~~~~~~\nonumber\\
&&+\tilde{Q}_{2}\exp\left[ -\frac{3}{2}\left(3-b-2\tilde{b}-\sqrt{\zeta}\right)x\right] ,\\  
\end{eqnarray}
where 
\begin{equation}
Q_{2}=\rho^{1}_{eff}=\rho_{1}-\dfrac{3}{4}\theta_{1},~~~~\tilde{Q}_{2}=\tilde{\rho}^{1}_{eff}=\tilde{\rho}_{1}-\dfrac{3}{4}\tilde{\theta}_{1},
\end{equation}
\end{subequations}
When $\zeta=0$ or $b_{\pm}=-1\pm 2\sqrt{2\tilde{b}}-2\tilde{b}$, and  looking only for solutions leading to an accelerated expansion (which corresponds here to the value $b_{+}=-1+2\sqrt{2\tilde{b}}-2\tilde{b}$ ), $\rho_{eff}$ and $\omega_{eff}$ become 
\begin{equation}
\rho_{eff}=\left(Q_{2}+\tilde{Q}_{2} \right)e^{-3\left(2-\sqrt{2\tilde{b}} \right)x}~~\mbox{and}~~\omega_{eff}=1-\sqrt{2\tilde{b}}.
\end{equation}
$\rho_{eff}$ decreases for $x\le 0$ and acceleration then occurs if $8/9<\tilde{b}<2$. On the other hand, $\rho_{eff}$ reduces to the cosmological constant with $\omega_{eff}=\omega_{\textsc{cc}}=-1$ for $\tilde{b}=2$.\\
~~~~~~~~~~~~~~~~~~~\\
Now, taking $\zeta\ne 0$ and assuming furthermore a monotonically decreasing effective density during the evolution of the universe the coupling $b$ and $\tilde{b}$ have to satisfy one of the following constraints:
\begin{equation}
\label{con}
 \begin{array}{rlll}
\alpha_{1}) &&\tilde{b}< 0 \Longrightarrow  b<1-\tilde{b},\\  
\alpha_{2})&& 0<\tilde{b}\leq 2 \Longrightarrow  b\leq -1-2\sqrt{2\tilde{b}}-2\tilde{b}~~~~~~~~~~~~~\\

 &&~~~~~~~~~~~~~~\mbox{or}~-1+2\sqrt{2\tilde{b}}-2\tilde{b}\leq b<1-\tilde{b},\\ 
\alpha_{3})&& \tilde{b}>2 \Longrightarrow  b\leq -1-2\sqrt{2\tilde{b}}-2\tilde{b}, 
\end{array}
\end{equation}
where we considered both positive and negative values for $b$ and $\tilde{b}$, and assumed that the integrations constants $Q_{1}$ and $\tilde{Q}_{1}$ are all nonzero and positive. The effective equation of state may be written as
\begin{equation}
\label{specc}
\omega_{eff}(x)=\dfrac{1}{2}\left[1-b-2\tilde{b}+\sqrt{\zeta}~\dfrac{1-\chi e^{3\sqrt{\zeta}x}}{1+\chi e^{3\sqrt{\zeta}x}} \right], 
\end{equation}
with $\chi=\dfrac{\varpi_{2}}{1-\varpi_{2}}$ and $\varpi_{2}=\dfrac{\tilde{Q}_{2}}{3H_{0}^{2}}$. 
Taking $b=\tilde{b}=1/3$ we obtain $\omega_{eff}^{-\infty}=\sqrt{3}/3$,~ $\omega_{eff}^{+\infty}=-\sqrt{3}/3<-1/3$ and $\omega_{eff}^{0}=(1-2\varpi_{2})\sqrt{3}/3$ which is less than $-1/3$ if $\varpi_{2}>1/2+\sqrt{2}/6\approx 0.78$.
The Figure (\ref{fig31.eps}) shows the plot of $\omega_{eff}$ in (\ref{specc}) versus $x$ for different values of $(b, \tilde{b}, \chi)$:  $b=1/3,~ \tilde{b}=1/3,~ \chi=9$ (dashed curve, $\omega_{eff}\approx -0.46$);~ $b=0.66,~ \tilde{b}=1/3,~ \chi=4$ (dash-dotted curve, $\omega_{eff}\approx -0.66$) and $b=0.66,~ \tilde{b}=1/3,~ \chi=9$ (solid curve, $\omega_{eff}\approx -0.83$), and the dotted line represents $\omega_{acc}=-1/3$.

Notice that in the special case where $Q_{2}=\tilde{Q}_{2}=\dfrac{q}{2}$, Eq. (\ref{special}) becomes
\begin{equation}
\rho_{eff}=q\cosh\left(\dfrac{3}{2}\sqrt{\zeta}x \right)\exp\left[ -\frac{3}{2}\left(3-b-2\tilde{b} \right)x\right]  
\end{equation}
which decreases monotonically when $x\le 0$. 

~~~~~~~~~~~~~~~~~~~~~~~~~~~~~~~~~~~~~~~~~\\
~~~~~~~~~~~~~~~~~~~~~~~\\
~~~~~~~~~~~~~~~~~~~~~~~~~~~~~~~~~~~~~~~~~\\
~~~~~~~~~~~~~~~~~~~~~~~\\

\subsection{Concluding Remarks}
 We have constructed, in normal gauge for a four-dimensional Lyra manifold, a dark energy model containing only the standard matter that interacts with the vector field $\phi_{\mu}$. Assuming a constant displacement vector field, the model mimics the $\Lambda$CDM model, indicating that the constant displacement field, considered as an energy component of the total energy, plays the role of a cosmological constant with $\omega_{\textsc{cc}}=\omega_{\phi_{int}}=-1$. Introduced in general relativity in an ad hoc fashion,  the cosmological constant arises in Lyra's geometry as a result of the presence of a vector field in the affine structure and has then an intrinsic geometrical significance. On the other hand, considering a time-dependent displacement vector field mutually interacting with the pressureless matter and without an alien energy component with a negative pressure, we have alleviated the fine-tuning and coincidence puzzles and shown that the universe could recently enter an accelerating phase, with even an equation of state crossing the cosmological constant boundary $\omega_{\textsc{cc}}=-1$. In fact, the geometrical contribution to the effective energy density, resulting from the modification of the Riemann manifold is endowed with a negative pressure that counteracts the gravitational forces produced by the ordinary matter and therefore favours the acceleration of the universe. In contrast to several models that consider an alien energy component in the universe, this work ensures that dark energy responsible for the late time acceleration of the universe has an intrinsic geometrical origin.

\section*{Acknowledgements}
I would like to thank Dr. K. Ayenagbo for valuable discussions

\section*{References}
\providecommand{\href}[2]{#2}\begingroup\raggedright\endgroup


\begin{thebibliography}{10}
\bibitem{Riess}A. G. Riess et al., Astron. J. \textbf{116}, (1998) 1009.
\bibitem{Perlmutter}S. Perlmutter et al., Astrophys. J. \textbf{517}, (1999) 565.
\bibitem{Tegmark1}M. Tegmark et al., Phys. Rev. D \textbf{69} (2004) 103501.
\bibitem{Tegmark2}M. Tegmark et al., Astrophys. J. \textbf{606} (2004) 702.
\bibitem{Tegmark3}M. Tegmark et al., Phys. Rev. D \textbf{74} (2006) 123507.
\bibitem{Seljak}U. Seljak et al., Phys. Rev. D \textbf{71} (2005) 103515.
\bibitem{Adelman}J. K. Adelman-McCarthy et al., Astrophys. J. Suppl. 
    \textbf{162} (2006) 38.
\bibitem{Abazajian}K. Abazajian et al., Astron. J. \textbf{126} (2003) 2081;
    Astron. J. \textbf{128} (2004) 502 and Astron. J. \textbf{129} (2005) 1755.
\bibitem{Spergel1}D. N. Spergel et al., Astrophys. J. Suppl. \textbf{170} (2007) 377.
\bibitem{Page}L. Page et al., Astrophys. J. Suppl. \textbf{170} (2007) 335.
\bibitem{Hinshaw}G. Hinshaw et al., Astrophys. J. Suppl. \textbf{170} (2007) 288.
\bibitem{Jarosik}N. Jarosik et al., Astrophys. J. Suppl. \textbf{170} (2007) 263.
\bibitem{Weinberg}S. Weinberg, Rev. Mod. Phys. \textbf{61}, 1 (1989); V. Sahni and A. A. Starobinsky, Int. J. Mod. Phys. D \textbf{9}, 373 (2000); S.M. Carroll, Living Rev.Rel. \textbf{4}, 1 (2001); P.J.E. Peebles and B. Ratra, Rev. Mod. Phys. \textbf{75}, 559 (2003); T. Padmanabhan,
Phys. Rept. \textbf{380}, 235 (2003).
\bibitem{Copeland}E. J. Copeland, M. Sami and S. Tsujikawa, Int. J. Mod.Phys. D 
\textbf{15}, 1753 (2006).
\bibitem{Amendola1}L. Amendola and S. Tsujikawa, Dark energy, CUP, 2010.
\bibitem{Hawking}S. W. Hawking, Phys. Lett. B \textbf{134}(1984) 403.
\bibitem{Kachru}S. Kachru, M. B. Schulz and E. Silverstein, Phys. Rev. D 
\textbf{62} (2000) 045021.
\bibitem{Tye}S.-H. H. Tye and I.Wasserman, Phys. Rev. Lett. \textbf{86}(2001) 1682.
\bibitem{Yokoyama}J. Yokoyama, Phys. Rev. Lett. \textbf{88} (2002) 151302.
\bibitem{Mukohyama}S. Mukohyama and L. Randall, Phys. Rev. Lett. 
\textbf{92} (2004) 211302.
\bibitem{Kane}G. L. Kane, M. J. Perry and A. N. Zytkow, Phys. Lett. B 
\textbf{609} (2005) 7. 
\bibitem{Dolgov}A. D. Dolgov and F. R. Urban, Phys. Rev. D \textbf{77} (2008) 083503.
\bibitem{Ratra}B. Ratra and P. J. E. Peebles, Phys. Rev. D \textbf{37}, 3406 (1988); P. J. E. Peebles and B.Ratra, ApJL \textbf{325}, L17 (1988); C. Wetterich, Nucl. Phys. B \textbf{302}, 668 (1988); I. Zlatev, L. Wang and P. J. Steinhardt, Phys. Rev. Lett. \textbf{82}, 896 (1999).
\bibitem{Onemli}V. K. Onemli and R. P. Woodard, Class. Quant. Grav. \textbf{19}, 4607 (2002) [\href{http://arxiv.org/abs/gr-qc/0204065v2}{arXiv:gr-qc/0204065v2}]; V. K. Onemli and R. P. Woodard, Phys. Rev. D \textbf{70}, 107301 (2004) [\href{http://arxiv.org/abs/gr-qc/0406098v2}{arXiv:gr-qc/0406098v2}]; T. Brunier, V. K. Onemli and R. P. Woodard, Class. Quant. Grav. \textbf{22}, 59-84 (2005) [\href{http://arxiv.org/abs/gr-qc/0408080v2}{arXiv:gr-qc/0408080v2}]; E. O. Kahya and V. K. Onemli, Phys. Rev. D \textbf{76}, 043512 (2007) [\href{http://arxiv.org/abs/gr-qc/0612026v3}{arXiv:gr-qc/0612026v3}]; E. O. Kahya, V. K. Onemli and R. P. Woodard, Phys. Rev. D \textbf{81}, 023508 (2010) [\href{http://arxiv.org/abs/0904.4811v1}{arXiv:0904.4811v1}].




\bibitem{Caldwell}R. R. Caldwell, Phys. Lett. B \textbf{545}, 23 (2002); S.M. Carroll, M. Hoffman and M. Trodden, Phys. Rev. D \textbf{68}, 023509 (2003).
\bibitem{Armendariz}C. Armend\'{a}riz-Pic\'{o}n, T. Damour and V. Mukhanov, Phys. Lett. B \textbf{458}, 209 (1999); C. Armend\'{a}riz-Pic\'{o}n, V. Mukhanov and P.J. Steinhardt, Phys. Rev. D \textbf{63}, 103510 (2001); T. Chiba, T. Okabe and M. Yamaguchi, Phys. Rev. D \textbf{62}, 023511
(2000).
\bibitem{Padmanabhan}T. Padmanabhan, Phys. Rev. D \textbf{66}, 021301(R) (2002); J. S. Bagla, H. K. Jassal, and T. Padmanabhan, Phys. Rev. D \textbf{67}, 063504 (2003).
\bibitem{Li}M. Li, Phys. Lett. B \textbf{603}, 1 (2004); Q. G. Huang and M. Li, JCAP \textbf{0408}, 013 (2004); Q. G. Huang and M. Li, JCAP \textbf{0503}, 001 (2005);  Q. G. Huang and Y. G. Gong, JCAP \textbf{0408}, 006 (2004);  X. Zhang and F. Q. Wu, Phys. Rev. D \textbf{72}, 043524
(2005); M. Li, X. D. Li, S. Wang and X. Zhang, JCAP \textbf{0906}, 036 (2009); Y. T. Wang and L. X. Xu, Phys. Rev. D \textbf{81}, 083523 (2010).
\bibitem{Limiao}M. Li et al., JCAP \textbf{0912}, 014, 2009. \href{http://arxiv.org/abs/0910.3855v2}{arXiv:0910.3855v2}.
\bibitem{Cai}R. G. Cai, Phys. Lett. B \textbf{657}, 228 (2007); H. Wei and R. G. Cai, Phys. Lett. B \textbf{660}, 113 (2008).
\bibitem{Wei}H. Wei, R. G. Cai, and D. F. Zeng, Class. Quant. Grav. \textbf{22}, 3189 (2005); H. Wei, and R.G. Cai, Phys. Rev. D \textbf{72}, 123507 (2005).
\bibitem{Kamenshchik}A. Y. Kamenshchik, U. Moschella and V. Pasquier, Phys. Lett. B \textbf{511}, 265 (2001); M. C. Bento, O. Bertolami and A.  A. Sen, Phys. Rev. D \textbf{66}, 043507 (2002); X. Zhang, F. Q. Wu and J. Zhang, JCAP \textbf{0601}, 003 (2006).
\bibitem{Zhang}W. Zhao and Y. Zhang, Phys. Lett. B \textbf{640}, 69-73 (2006); W. Zhao and Y. Zhang, Class. Quant. Grav. \textbf{23}, 3405-3418 (2006); Y. Zhang, T. Y. Xia, and W. Zhao, Class. Quant. Grav. \textbf{24}, 3309 (2007); T. Y. Xia and Y. Zhang, Phys. Lett. B \textbf{656}, 19 (2007); S. Wang, Y. Zhang and T. Y. Xia, JCAP \textbf{10}, 037 (2008); S. Wang and Y. Zhang, Phys. Lett. B \textbf{669}, 201 (2008).
\bibitem{Kerner}R. Kerner, Gen. Rel. Grav. \textbf{14}, No. 5, 453-469 (1982).


\bibitem{Capozziello}S. Capozziello, Int. J. Mod. Phys. D \textbf{11}, 483-492 (2002) [\href{http://arxiv.org/abs/gr-qc/0201033v1}{arXiv:gr-qc/0201033v1}].


\bibitem{Nojiri}S. Nojiri, S.D. Odintsov, Phys. Lett. B \textbf{576}, 5 (2003); S. Nojiri, S. D. Odintsov, Mod. Phys. Lett. A \textbf{19}, 627 (2003); S. Nojiri, S. D. Odintsov, Phys. Rev. D, \textbf{68}, 12352 (2003); S. M. Carroll, V. Duvvuri, M. Trodden, M. Turner, Phys. Rev. D \textbf{70}, 043528 (2004). 
\bibitem{Capozziello1}S. Capozziello, S. Carloni, A. Troisi, Rec. Res. Dev. Astron. Astrophys. \textbf{1}, 625 (2003) [\href{http://arxiv.org/abs/astro-ph/0303041v1}{astro-ph/0303041}].
\bibitem{Nojiri1}S. Nojiri, S. D. Odintsov, Gen. Rel. Grav. \textbf{36}, 1765 (2004); X. H. Meng, P. Wang, Phys. Lett. B \textbf{584}, 1 (2004).

\bibitem{Lyra}G. Lyra, Math. Z. \textbf{54}, 52 (1951).
\bibitem{Sen1}D. K. Sen, Z. Physik, \textbf{149}, 311 (1957). 
\bibitem{Sen2}D. K. Sen, Can. Math. Bull. \textbf{3}, 255 (1960).
\bibitem{Weyl}H. Weyl, Sitzber. Preuss. Akad. Wiss., 465-480 (1918).
\bibitem{Sen3}D. K. Sen and K. A. Dunn, J. Math. Phys. \textbf{12}, 578 (1971)
\bibitem{Sen4}D. K. Sen and J. R. Vanstone, J. Math. Phys. \textbf{13}, 990 (1972).
\bibitem{Scheibe}E. Scheibe, Math. Z. \textbf{57}, No. 1,  (1952) 65-74.
\bibitem{Halford}W. D. Halford, Aust. J. Phys. \textbf{23}, 863 (1970).
\bibitem{Halford1}W. D. Halford, J. Math. Phys. \textbf{13}, 1699 (1972). 
\bibitem{Soleng}H. H. Soleng, Class. Quant. Grav. \textbf{5}, (1988) 1489-1500 .
\bibitem{Soleng1}H. H. Soleng, Gen. Rel. Grav. \textbf{19}, No. 12,  1213 (1987). 
\bibitem{Hudgin}R. H. Hudgin, J. Math. Phys. \textbf{14}, No. 12, (1973) 1794-1799.
\bibitem{Beesham}A. Beesham, Aust. J. Phys. \textbf{41}, 833 (1988).
\bibitem{Beesham1}A. Beesham, Astrophys. Space Sci. \textbf{127}, 189 (1986)
\bibitem{Manoukian}E. B. Manoukian, Phys. Rev. D \textbf{5}, 2915-2922 (1972).
\bibitem{Matyjasek}J. Matyjasek, Astrophys. Space Sci., \textbf{207}, 313 (1993).  
\bibitem{Anirudh}A. Pradhan and A. K. Vishwakarma, J. Geom. Phys., \textbf{49}, (2004) 332-342.
\bibitem{Mohanty}G. Mohanty, G. C.Samanta, K. L. Mahanta, Theoret. Appl. Mech., \textbf{36}, No. 2, (2009)  157-166; G. Mohanty, K. L. Mahanta, R. R. Sahoo, Astrophys. Space Sci. \textbf{306}, (2006) 269; G. Mohanty, K. L. Mahanta, B. K. Bishi, Astrophys. Space Sci. \textbf{310}, (2007) 273; G. Mohanty, K. L. Mahanta, R. R. Sahoo: Communications in Physics, \textbf{17}, No. 3 (2007).
\bibitem{Rahaman}F. Rahaman, Fizika B \textbf{11}, (2002) 223; F. Rahaman, S. Chakraborty, S. Das,  N. Begum, M. Hossain, J. Bera, Pramana-J. Phys. \textbf{60}, (2003) 453;  F. Rahaman, S. Das, N. Begum, M. Hossain,  Pramana-J Phys. \textbf{61}, (2003) 153.
\bibitem{Katore}S. D. Katore, S. V. Thakare and K. S. Adhao, Pramana-J. Phys. \textbf{71}, No. 1, (2008) 15-22. 
\bibitem{Gad}R. M. Gad, Can. J. Phys. \textbf{89}, 773–778 (2011).
\bibitem{Shchigolev}V. K. Shchigolev and E. A. Semenova,  [\href{http://arxiv.org/abs/1203.0917v1}{arXiv:1203.0917v1}].


\bibitem{Kunz}P. S. Corasaniti, M. Kunz, D. Parkinson, E. J. Copeland and B. A. Bassett, Phys. Rev. D\textbf{70} (2004) 083006.
\bibitem{Alam}U. Alam, V. Sahni, T. D. Saini and A. A. Starobinsky, Mon. Not. Roy. Astron. Soc. \textbf{354}, 275 (2004).
\bibitem{Guo}Z. K. Guo et al., Phys. Lett. B \textbf{608}, 177 (2005); J.-Q. Xia, B. Feng, and X. Zhang, Mod. Phys. Lett.  A \textbf{20}, 2409
(2005); M.-Z Li, B. Feng, and X.-M Zhang, J. Cosmol. Astropart. Phys. 12 (2005) 002; B. Feng, M. Li, Y.-S. Piao and X. Zhang, Phys. Lett. B \textbf{634}, 101 (2006); M. R. Setare, Phys. Lett. B \textbf{641}, 130 (2006); M. R. Setare and E. N. Saridakis, Phys. Rev. D \textbf{79}  043005 (2009); W. Zhao and Y. Zhang, Phys. Rev. D \textbf{73}, 123509 (2006); G.-B. Zhao, J.-Q. Xia, B. Feng, and X. Zhang, Int. J. Mod. Phys. D \textbf{16}, 1229 (2007); M. R. Setare, J. Sadeghi, and A. R. Amani, Phys. Lett. B \textbf{660}, 299 (2008); J. Sadeghi, M. R. Setare, A. Banijamali and F. Milani, Phys. Lett. B \textbf{662}, 92 (2008); M. R. Setare and E. N. Saridakis, Phys. Lett. B \textbf{668}, 177 (2008); M. R.
Setare and E. N. Saridakis, Int. J. Mod. Phys. D \textbf{18}, 549-557 (2009) [\href{http://arxiv.org/abs/0807.3807v1}{arXiv:0807.3807v1}]; M. R. Setare and E. N. Saridakis, J. Cosmol. Astropart. Phys. 09 (2008) 026.




\end{thebibliography}
\end{document}